# K model for designing Data Driven Test Automation Frameworks and its Design Architecture "Snow Leopard"

Rohan R. Kachewar
Automation Architect

## ABSTRACT

Automated testing improves the efficiency of testing practice at various levels of projects in the organization. Unfortunately, we do not have a common architecture or common standards for designing frameworks across different test levels, projects and test tools which can assist developers, testers and business analysts.

To address the above problem, in this paper, I have first proposed a unique reference model and then a design architecture using the proposed model for designing any Data Driven Automation Frameworks. The reference model is "K model" which can be used for modeling any data-driven automation framework. The unique Design architecture, based on above model is "Snow Leopard".

## 1. INTRODUCTION

A test automation framework is a set of assumptions, concepts and tools that provide support for automated software testing. Historically, test automation has not met with the level of success that it could. Time and again test automation efforts are born, stumble, and die. Most often this is the result of incorrect design, lack of flexibility for future enhancements, unchecked redundancy etc.

Automation Frameworks can be generally classified as follows:

1. Data driven
2. Keyword driven
3. Modular
4. Hybrid: A combination of above three.

Data Driven Automation Frameworks are used generally for applications requiring fixed set of actions to be performed, but with a lot of permutations and combinations of the various parameters which form the test cases.

The key requirements of a data driven automation framework would be as follows:

- A user friendly UI [User Interface] to interact and use the framework, so as to hide the complexities of the test framework.
- A generic algorithm to reduce the huge number of test cases that can arise from all the possible permutations

and combinations of parameters and their values in the application.
- An easy way to create test cases without need of any programming involved.
- An application specific Log generation module.
- A nightly report generation and mail module, for sending the results of suites run
- Lastly, use of effective techniques to reduce the Log analyzing and Bug reporting time.

Although there is a significant body of existing research on software test automation, a review of the literature reveals there are mostly too specific to a particular application. Also none discuss the actual problems faced in framework development and efficient solutions to those.

## 2. LITERATURE REVIEW

I have undergone literature review phase and evolved with a problem statement with the help of work that has been published till today in the area of automation framework design for Data Driven Frameworks.

Most published research on test automation frameworks presents case studies or feasibility studies.

A 2011 paper by Artzi et al, describes rudimentary test automation framework for perform feedback-directed test generation for JavaScript web applications [1]. That paper presented the case study of a specific system (rather than general or all web applications) also it needed access to AUT's source code.

Another 2011 paper by James M. Slack described a test automation framework using "AutoIt" tool and an Excel Sheet as a Data Container for test data [2]. The framework proposed was tool specific and the disadvantages of using excel sheet as a data container were not covered, as excel sheet representation is not suitable for more complex and dynamic web applications.

A 2010 paper by Manpreet Kaur et all "Xml Schema Based Approach for Testing of Software Components" mentions apt use of XML format for representing test data [3].

Another 2010 paper "DEVELOPMENT OF TEST AUTOMATION FRAMEWORK FOR TESTING AVIONICS SYSTEMS" describes aptly some basics for implementing data driven frameworks but again it does not give a generic model or architecture for general design. [4]

A 2002 paper by Tsai, Paul, Song, and Cao presented a description of an XML based framework named Coyote which was designed for testing Web services [5]. Again, this paper was a case study and presented no conclusions.

A 2009 paper by Merchant, Tellez, and Venkatesan presented a browser agnostic UI test framework for Web applications and concluded that using the framework reduced the time required to create test case scenarios by 50% compared to a manual approach [6].

## 3. K MODEL:

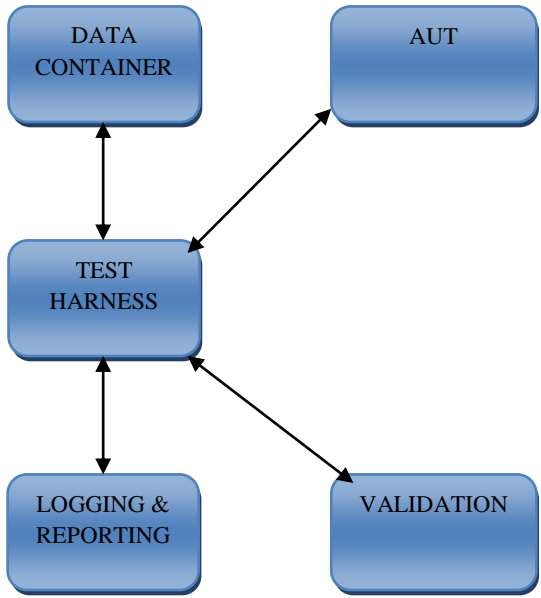

**Figure 1: K Model**

**K Model:**
The model encapsulates all complex components of a data driven test framework into 5 major modules.
This can be treated as black boxes for designing and further expanded upon as:

- **Test Harness:** the heart of any framework, it contains the run-test or test driver, suites, and test library.
- **Data Container:** The module containing the test data, for the data driven framework.
- **AUT Driver:** [Application under Test Driver] Mostly data driven frameworks are for testing Web applications or GUI applications, so the AUT driver is a module which actually does all the actions as per the test case on the application, i.e. actual interaction with the application.
- **Validation:** This module is used to design the components for validating whether the configuration done using AUT Driver is correct or not.
- **Log Generation and Reporting:** This module will be responsible for generation of test case execution logs and the Report Generation for all the test suites executed.

## 4. SOLUTION ARCHITECTURE: SNOW LEOPARD

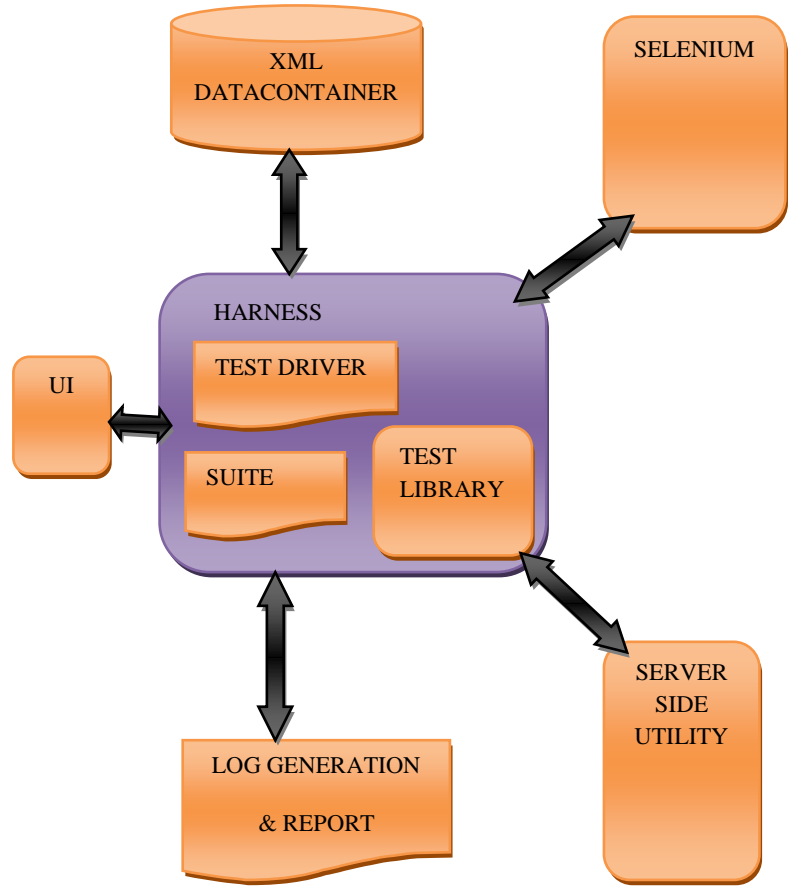

**Figure 2: Solution Architecture: Snow Leopard**

### 4.1 Introduction:

Snow Leopard is a general architecture which can be used as a template for designing K model based Automation Test Frameworks. Its successful implementation in Perl and Selenium, is explained further, however one can implement "Snow Leopard" architecture using any other language and tool combination as well. The proposed architecture uses XML format to store test data and a unique test case reduction algorithm, which is an important concern for huge number of test cases in data driven frameworks. Furthermore, detailed description about usage of XML to express test cases and usage of a GUI template for ease of framework maintenance is also explained. The frame work was build using Perl so I have provided references accordingly, however the same concepts can be modeled for other scripting languages as well.
The architecture contains 6 modules each of which is designed and explained in detail further.

Proposed Directory Structure for the framework:

- **Bin:** contains all executables and binaries,
  Example: run-test, log cleanup scripts etc.
- **Test Case:** Contains all test case scripts.
- **Test Suite:** A File containing all test cases to be run.
- **Lib:** Contains all Base-class and library files.
- **Logs**: Contains all test case execution logs
- **Utils:** Contains all 3[rd] party utilities required.

## 4.2 UI [User Interface]:

Most of the existing frameworks architectures do not have a User friendly UI to manage the framework, but are command line driven, making it difficult to use, for new people. Also lack of UI exposes the framework to threat of incorrect usage which may cause further failure.

**Proposed Design:** Any Automation Framework UI must contain following functionalities:
- Dynamic Suite creation facility
- Facility to Schedule Suite Runs
- Facility to run  Individual Test case
- Build Selection
- Dashboard

**Implementation**: I had implemented the UI using Perl-Tk Applet programming. Perl's Tk module [7] facilitates a decent and rapid UI creation. The Dynamic suite creation was implemented by 2 ways:
- Multiple test case selection by use of check boxes in front of all the test case names, to identify which cases to run.
- However when cases are in large numbers like 100-200, such multiple selection is tedious and so an excel sheet which was the, actual suite file had a column in front of the test case name  the "Run" Column which enabled/disabled the test case run decision. As shown in Figure 2. Also the facility to run suite as per the priority is also necessary.
  **Bats:** Basic Acceptance tests, to be run on every build
  **P1:** Less priority than Bat and similarly cases can be distinguished as P2, P3 priority as well.

| Test case | Run | Priority |
|-----------|-----|----------|
| TC1 | Y | Bat |
| TC2 | N | P1 |

**Figure 3: Test Suite representation using Excel Sheet**

- **Scheduling:** This is required especially when one has to run test suites overnight.
  **Implementation:** In the UI when time to start a selected suite was entered, internally it would schedule the job using simple "AT" command of windows. Similar commands as per the OS can be used, example the cron tab in UNIX.

- **Build Selection**: As the build installation procedure was automated, merely providing the build number would download and install the build before starting the suite execution.
  **Implementation:** One generally needs to ftp the build from a server and then install, this was achieved by modules like Net::FTP in Perl.

- Lastly the **Dashboard** section was more for the management to check the testing results graphically.
  **Implementation:** There are many freeware graph generators available, also modules in Perl; however I had used the "Google Charts", which generates a graph from the queried url itself, that too in required image format. So for example if someone wants to check the testing status for June 2011 month delivery, internally a Perl script will parse the excel Report of June 2011 and get the total, passed and failed number of test cases and accordingly generate a url to generate a pie chart using Google Charts [8]. There is also a Perl module Google::Chart for the same.

## 4.3 Harness:

### 4.3.1 Run-test/Test Driver:

This is one of the most important elements of any framework. It is a driver utility which actually executes all the test cases.
**Proposed Design:** A good harness or run-test module must facilitate following functionalities:
- Install  the build as selected using UI
- Run all the applicable test cases in the scheduled suite.
- Fault/Crash handling mechanism
- Give the  entire Suite Run Summary
- Interact with the Report module.
- Initiate the Mail module at end of the entire suite run.

**Implementation:**
- First the build installation would be done.
- Then harness would parse the Excel Suite file as per the set instructions and generate an array of test case ids to be executed. These would then be executed one by one by passing the corresponding Perl script to the Perl interpreter.
- The most important aspect of the harness is to handle the test case failures properly. In-case there is a Crash or abnormal failure in a test case, and the failure is not handled properly, it may cause the entire suite to be aborted. This was taken care by a setup and cleanup methods at both test case as well as suite level. Also use of "try and catch" in Perl further avoids such issues. In case of crashes, the test case cleanup would identify such an issue and inform the harness at the end of test case execution after saving necessary "crash dump" in clean up phase. The harness would immediately reload the host machine so as to regain the original working environment as before. Also default timeout monitoring at harness level for each running test case, helps avoiding test cases going into infinite loops scenarios.
- The entire Suite run Summary as the number of cases failed passed and total executed should be calculated

in at harness level. A good example of such functionality is the Perl's Test::Harness module.

- The interaction with Report module: The harness would create an Excel sheet at beginning of every suite run with fields as shown in Figure 5. And as soon as a test case execution finishes it would fill the corresponding entries in the report for that test case. In cleanup of the test case if the case failed, the harness would query the database of our Bug management tool, with "test case id" as primary key and update the Report with the known open bug id, if any against that test case.

- Finally at the end of suite run harness would initiate the mail module.

### 4.3.2  Suites

These should contain the grouping of test cases as per features to be tested.

**Proposed Design:**

- Every major feature should have a dedicated Suite File; this facilitates tracking of testing progress and defects for any release with respect to each feature.
- Each Suite should contain test cases grouped as per priority i.e. BAT, P1, P2.
- Whether a test case should be run or not should be indicated from the suite file.

**Implementation:**

 Please refer figure 3, for test suite implementation.

We had used a simple excel file as a suite file and the Perl module "Spreadsheet::ParseExcel" for parsing the suite files.

### 4.3.3  Library Module:

This is a centralized repository of all the required Base classes or parent Classes which contain all the reusable methods and data variables.  Designing this module is a very important aspect as in most of the failed automation experiments; the key reason of failure was inflexible design of base classes causing bulky and complicated inheritance pattern and lot of redundancy.

**Design Proposed:**

- Keep inheritance as less deep as possible, ideally it should not me more than a level of 3 [i.e. Grandparent->Parent->Child]
- Usage of hash data structure while interacting with base class methods.
- Avoid adding redundant methods as far as possible.
- Separate Base classes should be there for each major functionality.

**Implementation:**

- Deep inheritance makes debugging very difficult in case of failures, so we had kept inheritance less deep as far as possible.
- Usage of hash to pass arguments is a very important concern. As the Application under Test grows old, more and more new features get added. Thus in new test cases, to add support for older Base-Class methods, become difficult. If one goes on passing scalar data variables as flags, to indicate handling as per new cases, the method calling becomes bulky and complicated and also one has to change the previous method calls made in older cases so as to avoid syntax errors in compilation. So use of hash encapsulates the flags and makes the code more readable and simpler.

Example: if earlier a method addPlayer was used for generating a player of given name but default frequency.

i.e. addPlayer($name)

Now as per a new feature I can also configure frequency and so now the support added in base class for new feature, requires method call as **addPlayer( $name, $freq)**.

Thus now I have to go back and change the old test cases and give a default frequency variable in method call, to avoid compilation errors when running old cases.

However the same, if I had kept a hash data structure in beginning:

**%hash = {"name" => abc}**
**addPlayer(\%hash)**

Thus for new feature support I just have to add another key to the hash and set a default value in method. So no need of changing the earlier method calls in old test cases either.

**%hash = {"name" => abc, "freq" => xyz}**
**addPlayer(\%hash)**

Such flexibility in Base-classes avoids a lot of future rework.

## 4.4  XML Data Container and Test Case:

### 4.4.1  XML Data Container:

This module is the brains of any data driven framework. It contains all the environment test data for execution of any test case. Some people also prefer excel sheet as a data container, but this becomes a problem as the framework begins to grow, and also excel sheet solution is not efficient for more complex application under test.

**Design Proposed:**

- Proper representation of structure so as to support varied user defined tags to describe content.
- Should contain a Primary key so as to differentiate data variables of one test case from another.
- Tags used should contain meaningful names.
- Avoid deep XML patterns.
- The XML parser should convert the XML file into a nested Hash Data structure.

**Implementation:**

For using a XML as data storage, we needed an efficient XML parser. We used the "XML::Simple" module of Perl. This converted the file into a nested Hash data structure and hence accessing data variables was much easier. Refer figure 4 and 5.

```
<tcs >
    <tc name="tc1" playername="man1" freq="50 or 25">
    </tc>
    <tc name="tc2" playername="man2" freq="29.97">
    </tc>
</tcs>
```

**Figure 4.1: XML format for data container**

Hash formed after parsing by XML parser:
```
{
    "tc" => {
            "tc1" => {
                        "playername" => "man1",
                            "freq" => "50 or 25",
                    };

            "tc2" => {
                        "playername" => "man2",
                            "freq" => "29.97",
                    };

            };
}
```
**Figure 4.2:  [Perl representation of a hash]**

Thus the test case id i.e. "tc1" or "tc2" was used as a primary key. Each test case had an independent hash with the parameters like player-name and frequency as keys and their corresponding data as values.

**Working**:

Each test case would tell the harness to parse the XML file as per the test case id and return the corresponding data to the test case, during the setup phase itself.

### 4.4.2  Test Case:

Test case is a script containing particular test scenario to be tested.

**Design Proposed:**

- Each test case should be an independent script.
- It should be structured generally as below:
  **Setup phase**: Initial environment set
  **Steps** 1, 2, 3 and so on.
  **Validation Phase**
  **Cleanup Phase:**

**Implementation:**

Each test case was a Perl script, e.g. "tc1.pl"

- The **Setup Phase** as mentioned: We parsed the XML and collected the test data.
- Since we had built the framework to test a Web Based Application used for configuring server side devices, in **Steps** we did configurations using application as per the test scenario. The tool used for this interaction with Web Application was Selenium, owing to its support to Perl platform.
- In **validation** Phase we cross checked that whether the configured actions on application are reflected on server side, by parsing the server side Logs for configured test data.
- Lastly in **cleanup phase** we deleted the made configurations so as to free the resources for next test case, and also closed the log file handle in setup.

## 4.5  AUT Driver [Selenium]:

The driver is generally a 3rd party tool which helps us to execute the manual actions by using an automation script. A major feature in such tools is the record and playback functionality [9].

**Design Proposed:**

- There are many freeware and licensed tools available. One should choose as per the requirement.
- If an option, choose a language for interacting with application, which is also useful for scripting at server side, for example Perl.
- Validate each and every page when browsing by automation
- Keep sufficient sleep interval in scripts, for allowing the page to load and also when any parameter is configured.

Implementation:

We used Selenium for automating the manual actions to be made on GUI of the web applications. Mostly owing to below reasons:

- Use of Selenium as it allows multi language platform support, like Perl, Python, and Java and so on.
- For Flex based application under test, a Flex plug-in for selenium is available, called as "Flex Monkium".

## 4.6  Server Side Utility [Validation Phase]:

The configuration done on web application should be verified at backend.

**Design Proposed:**

- Use separate library file for the validation phase.
- Many a times the value configured at application may appear differently while validating, in such cases we can add one more tag, in XML data container for value expected, so as to use it as reference while validating.

**Implementation:**

The development had given us a utility [or an executable] which would query the server side devices, (with arguments passed to utility by command line) and display the configuration in a file from which we parsed the key parameter values and cross checked with those in XML Data container.

In cases where the validation utility is a windows object, one can automate that using "Win32::OLE" Perl module and WinSpy++ Utility.

## 4.7  Logging and Report Generation:

This module encompasses the design for Log generation module and Report Generation Module and lastly the Mail module for mailing the result to interested parties.

**Design Proposed:**

- Logs should be application specific and detailed.
- There should be proper tags for each line in log file, like Info, Failed and Passed to facilitate debugging.
- There should be time stamp for each line of log.
- In case of failures the log should mention the failed test script file along with line number at which failure occurred.
- Final Report should contain, the suite name, and list all test cases with pass/fail status and also the failed cases should have a previously logged bug id in front of them, in case it was a known issue.

**Implementation:**

At suite setup:

- A folder with current time stamp as name was created and for all logs to be written in it.
- Also report file [excel sheet] was created within above folder.

Before starting every test case execution the run-test/test driver would make an entry in Report file of corresponding test case id and start time.

After every test case execution, the run-test/test driver would again update the Pass/Fail status, the end time and Bug id [got from querying the data base of Bug Management Tool for failed test cases] against every test case

In test case Setup phase itself we created a log file for each case and the same log file object was passed to Base class/Library files, as and how methods were called, so that all logs were written to same file handle for each test case. The Logger module would take care of this. This module was inherited in all files and had different methods as per the level of Log message, i.e. for "Info" level log message there was a separate method, and so was for "Failed" level messages. These facilitated adding of respective tags when called separately. [Refer the sample log below] In cleanup phase of test case the log file handle was closed.

Thus at end of the suite execution in Suite cleanup the report file handle was closed, and attached to outlook mail, and send to interested parties.

**Sample Log File:**
3-8-2010 12:07:28 [INFO] Logging Started...
3-8-2010 12:07:28 [INFO] Opening SM Application...
3-8-2010 12:07:42 [INFO] Loaded Config Page...
3-8-2010 12:07:44 [INFO] Clicked Player Config Link...
3-8-2010 12:08:25 [INFO] Loaded Player List Page...
3-8-2010 12:08:34 [FAILED] at C:\Perl\lib\Class.pm line 113.

**Sample Report:**

| ID | Start Time | End Time | Status | Bug id |
|----|-----------|----------|--------|--------|
| tc1 | 12:07 | 12:10 | Pass | |
| tc2 | 12:11 | 12:12 | Fail | PR2410 |

**Figure 5: Sample Report using Excel representation**

The Mail module can be implemented using the "MIME::Lite" module of perl.

# 5. TEST CASE REDUCTION ALGORITHM

The main issue in data driven frameworks is the huge number of test cases that can occur due to all possible permutations and combinations of the parameter and its possible values.

For instance, if you want to test with ten parameters of 26 values each, all combinations leads to 141,167,095,653,376 test cases.

In such a case neither is the test case execution possible nor is the test case writing. Imagine one person just writing all 141,167,095,653,376 test cases.

**Design Proposed:**

So in such a scenario we need some algorithm and automation which will:

- Take the parameter and expected values for each parameter from a text file and return me all the possible combinations in a Excel sheet as test cases. Thus automatic test case writing is achieved.

- Then we need an algorithm which will operate over all combinations and return only the pair-wise, triplet-wise …, combinations.

Concept behind such a strategy:

Refer the graph below [figure 6] reference [10]:

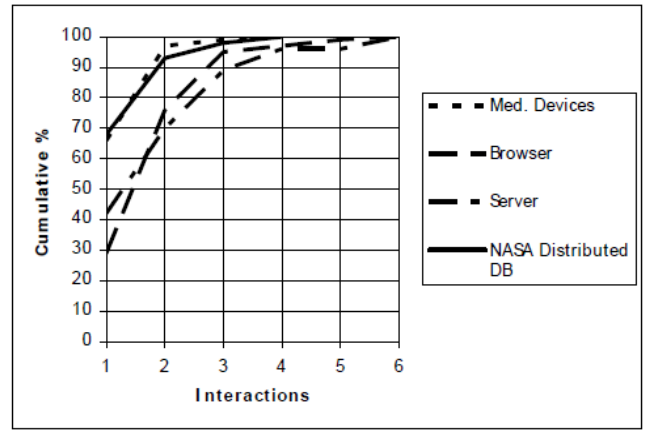

**Figure 6: Error-detection rates for four- to six way interactions in four application domains: medical devices, a Web browser, an HTTP server, and a NASA distributed database. Error detection rates for strength [interaction of variables] 1 to 6 v/s Cumulative defects found %.**

In the graph above the interactions between variables increase i.e. strength 2 indicates the testing done for all pair-wise combinations of parameters, whereas strength 3 denotes the testing done for all 3 way combinations of parameters. The cumulative % is the percentage of defects found for that particular combination testing. As seen the defect finding rate decreases beyond 2, i.e. the maximum defects can be found for all pair-wise combinations of parameters. Thus we can conclude that a satisfactory level of testing can be achieved if we cover all pair-wise combinations first.

**Implementation:**

- The all possible combinations can be achieved by simple nested for-loops implementation for all parameters; each for-loop iterating over all the values that parameter in that loop can take.

- The actual pair wise combinations can be achieved by free-ware Perl scripts available, for example the, AllPairs.pl [11] by James Bach is an ideal tool.
  For 3 way, 4 way and likewise combinations many licensed tools are present in the market.

# 6. CONCLUSION

The above proposed model i.e. K model facilitated us development of the actual framework in a properly planned and organized manner, rather than a haphazard one which would have been without any reference model. We first developed the design of each block in the K model and then the interfaces and interactions between each block.

The design proposed in Snow Leopard architecture, is an outcome of many years of experience in automation framework designing and hence helped, us avoid pit falls. The entire implementation helped us the cut down the manual efforts by to almost 1/4[th] i.e. 25 percent of that without the automation.